\documentclass[aps,pra,reprint,amsmath,amssymb,amsfonts,showkeys,showpacs,preprintnumbers]{revtex4-1}
\usepackage{bm,graphicx,xcolor,hyperref,rotating,lineno,subfigure,microtype}

\newcommand{\ha}{\frac{1}{2}}
\newcommand{\tha}{\frac{3}{2}}

\newcommand{\scL}{\mathcal{L}}

\begin{document}

\title{Harmonically trapped jellium}

\author{Pierre-Fran\c{c}ois Loos}
\thanks{Corresponding author}
\email{loos@rsc.anu.edu.au}
\author{Peter M. W. Gill}
\email{peter.gill@anu.edu.au}
\affiliation{
Research School of Chemistry, 
Australian National University, 
Canberra, ACT 0200, Australia}
\date{\today}

\begin{abstract}
We discuss the model of a $D$-dimensional confined electron gas in which the particles are trapped by a harmonic potential. In particular, we study the non-interacting kinetic and exchange energies of finite-size inhomogeneous systems, and compare the resulting Thomas-Fermi and Dirac coefficients  with various uniform electron gas paradigms. We show that, in the thermodynamic limit, the properties of this model are identical to those of the $D$-dimensional Fermi gas.
\end{abstract}

\maketitle

\section{Introduction}

Recent technical advances based on Bose-Einstein condensation in vapors of bosonic atoms \cite{Anderson95, Davis95, Bradley97, Fried98} have led to the experimental realization of ultracold Fermi gases composed of dilute gases of fermionic alkali atoms \cite{DeMarco99, Truscott01, Schreck01, Granade02, Roati02, Hadzibabic03}. These experiments are usually performed in harmonic traps using magneto-optical confinement techniques, and it is now possible to tune the harmonic trap to obtain not only three-dimensional gases but also quasi-two- and quasi-one-dimensional Fermi systems. Such experiments have been the driving force of numerous theoretical studies both at zero \cite{Butts97, Bruun98, Vignolo00, Vignolo03, Brack01, March01, Gleisberg00, Brack03, Mueller04} and finite \cite{vanZyl03a, vanZyl03b, Wang02} temperature.

The $D$-dimensional version of the jellium model (or $D$-jellium) consists of interacting electrons within an infinite volume and in the presence of a uniformly distributed background positive charge, and is the foundation of most density functionals. Traditionally, this system is constructed by allowing the number $n$ of electrons in a $D$-dimensional cube of volume $V$ to approach infinity with $\rho = n/V$ held constant \cite{Vignale,ParrYang}. 

A weakness of the electrons-in-a-box model is that it yields a uniform density only in the thermodynamic (i.e. $n\to\infty$) limit \cite{Jellium05}. We have recently \cite{Glomium} introduced an alternative model called $D$-spherium \footnote{This generalizes our earlier work \cite{Quasi09} in which ``$D$-spherium'' was a two-electron system \cite{EcLimit09, EcProof10, Frontiers10}.}, in which the electrons are confined to the surface of a $D$-sphere \footnote{We adopt the convention that a $D$-sphere is the surface of a ($D+1$)-dimensional ball}. This system possesses a uniform density, even for finite $n$, and because all the points in a $D$-sphere are equivalent, its mathematical analysis is relatively straightforward \cite{TEOAS09, Quasi09, LoosConcentric, LoosHook, LoosExcitSph}. In Ref.~\cite{Glomium}, we have shown that the properties of $D$-spherium can be calculated for finite $n$ and approach those of $D$-jellium as $n\to\infty$. 

In this paper, we will study the non-interacting kinetic and exchange energies of a spin-polarized many-electron system trapped in an isotropic harmonic trap \footnote{Anisotropy effects can be taken into account using the methodology developed in Ref.~\cite{Anisotropic}}. These quantities are of great importance in the framework of density-functional theory (DFT) \cite{ParrYang} for studying inhomogeneous systems and finite-size effects \cite{Kwee08, Ma11, UEGs}. We will compare the resulting Thomas-Fermi and Dirac coefficients with various uniform electron gas paradigms, such as the jellium and spherium models.

We will focus our attention on the physically important $D=2$ and $D=3$ systems. The $D=1$ case will be also studied due to the importance of the Bose-Fermi mapping in one-dimensional systems (bosonization) \cite{Girardeau60, Lieb63}.

\section{Trapped jellium model}

We consider a system of $n$ interacting electrons trapped in the $D$-dimensional isotropic harmonic potential 
\begin{equation}
	V(\bm{r}) = \ha r^2,
\end{equation}
where $r = |\bm{r}|$.  The Hamiltonian is 
\begin{equation}
	\Hat{H} = \sum_{i=1}^n \left[ -\frac{\nabla_i^2}{2} + V(r_i) \right]+ \sum_{i<j}^n \frac{1}{r_{ij}}
\end{equation}
with $r_{ij} = \left| \bm{r}_i - \bm{r}_j \right|$. 

The $\scL$th orbital of an electron in a harmonic trap is
\begin{equation}
	\Psi_\scL(\bm{r}) = \prod_{i=1}^{D} \psi_{\ell_i} (x_i),
\end{equation}
where $x_{i}$ is the $i$th cartesian coordinate of the electron. The composite index $\scL$ is given by 
\begin{equation}
	\scL = \left\{\ell_1,\ell_2,\ldots,\ell_D\right\},
\end{equation}
where the $\ell$'s are non-negative integers. The functions $\psi_{\ell}$, which satisfy the one-dimensional Schr\"odinger equation
\begin{equation}
\label{eigenfunctions}
	-\ha \frac{d^2 \psi_{\ell}}{dx^2} 
	+ \ha x^2 \psi_{\ell} = \epsilon_{\ell}\,\psi_{\ell},
\end{equation}
with $\epsilon_\ell = \ell + 1/2$, are the one-dimensional harmonic oscillator wave functions
\begin{equation}
	\psi_{\ell}(x) 
	= \frac{1}{\sqrt{2^{\ell} \ell! \pi^{1/2}}} 
	H_{\ell}(x) \exp\left(-\frac{x^2}{2}\right),
\end{equation}
where $H_{\ell}$ is the $\ell$th Hermite polynomial \cite{NISTbook}. We confine our attention to full-shell ferromagnetic systems, that is, every orbital with $ \ell_1+\dots+\ell_D \leq L$ is occupied by one spin-up electron. 

\begin{figure*}
	\subfigure[~$D=1$]{
	\includegraphics[width=0.45\textwidth]{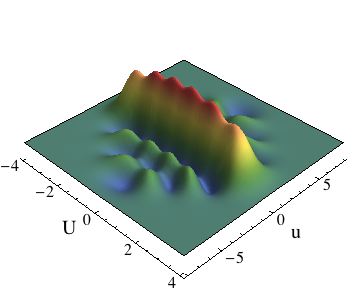}
	\includegraphics[width=0.45\textwidth]{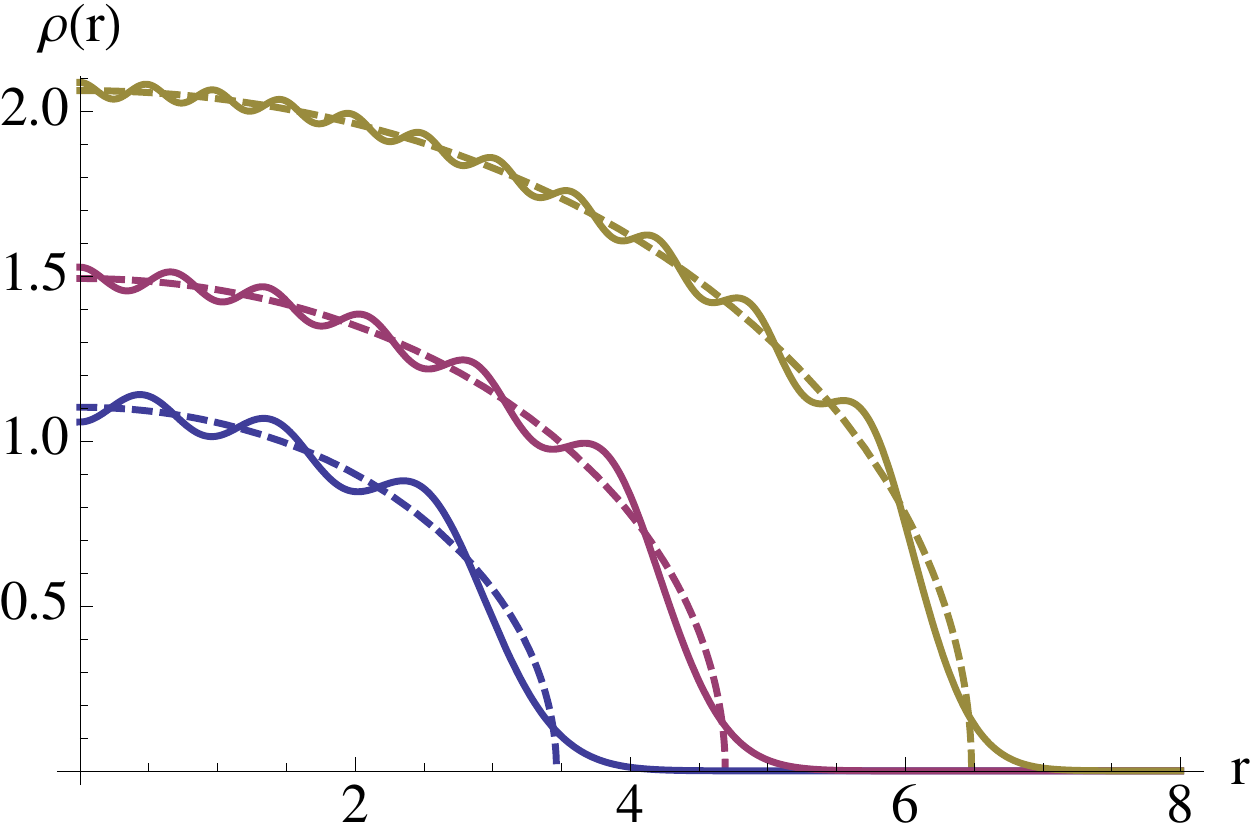}
	}
	\subfigure[~$D=2$]{
	\includegraphics[width=0.45\textwidth]{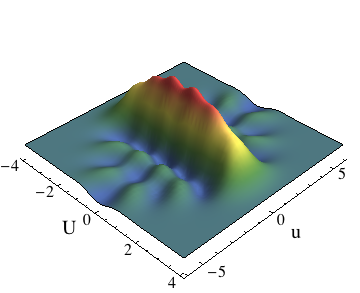}
	\includegraphics[width=0.45\textwidth]{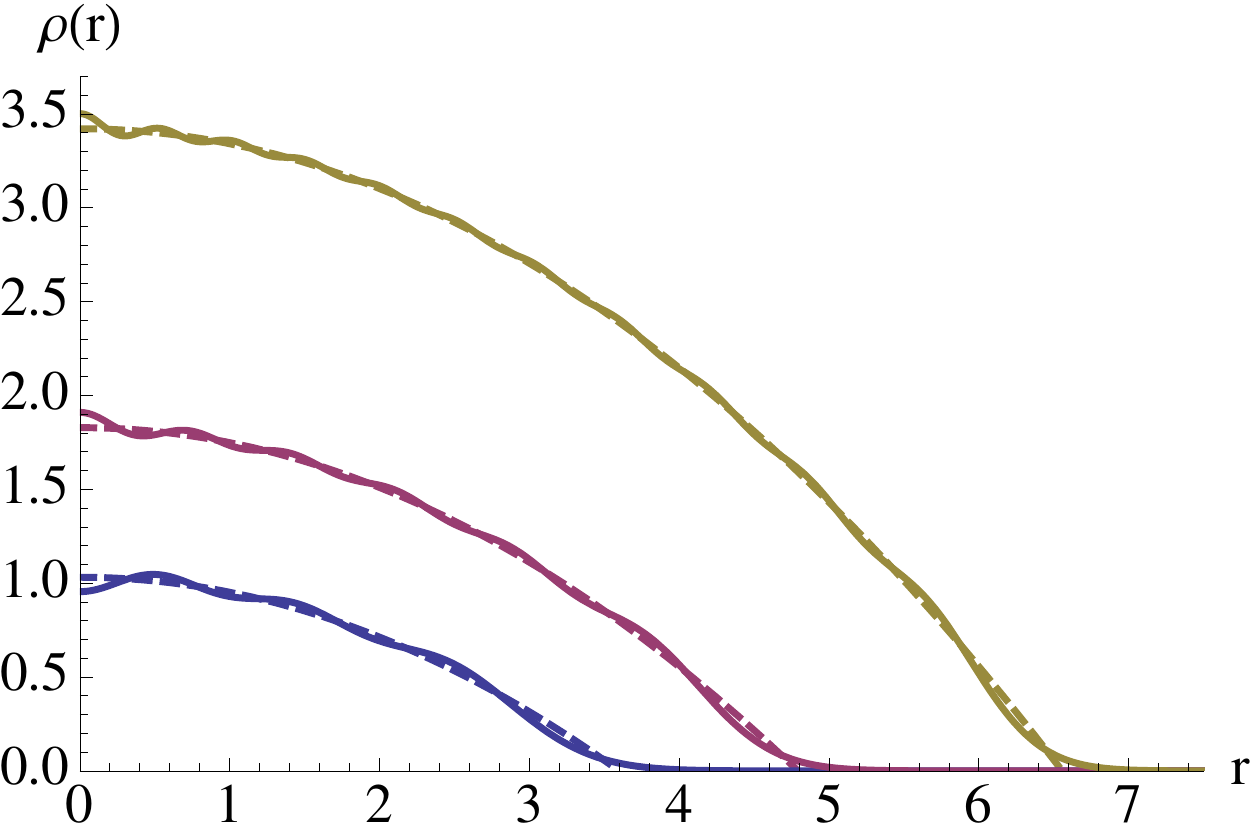}
	}
	\subfigure[~$D=3$]{
	\includegraphics[width=0.45\textwidth]{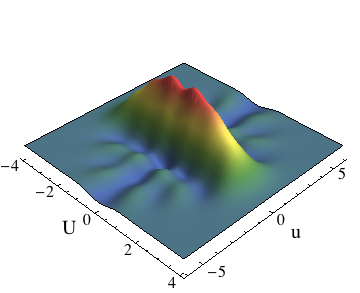}
	\includegraphics[width=0.45\textwidth]{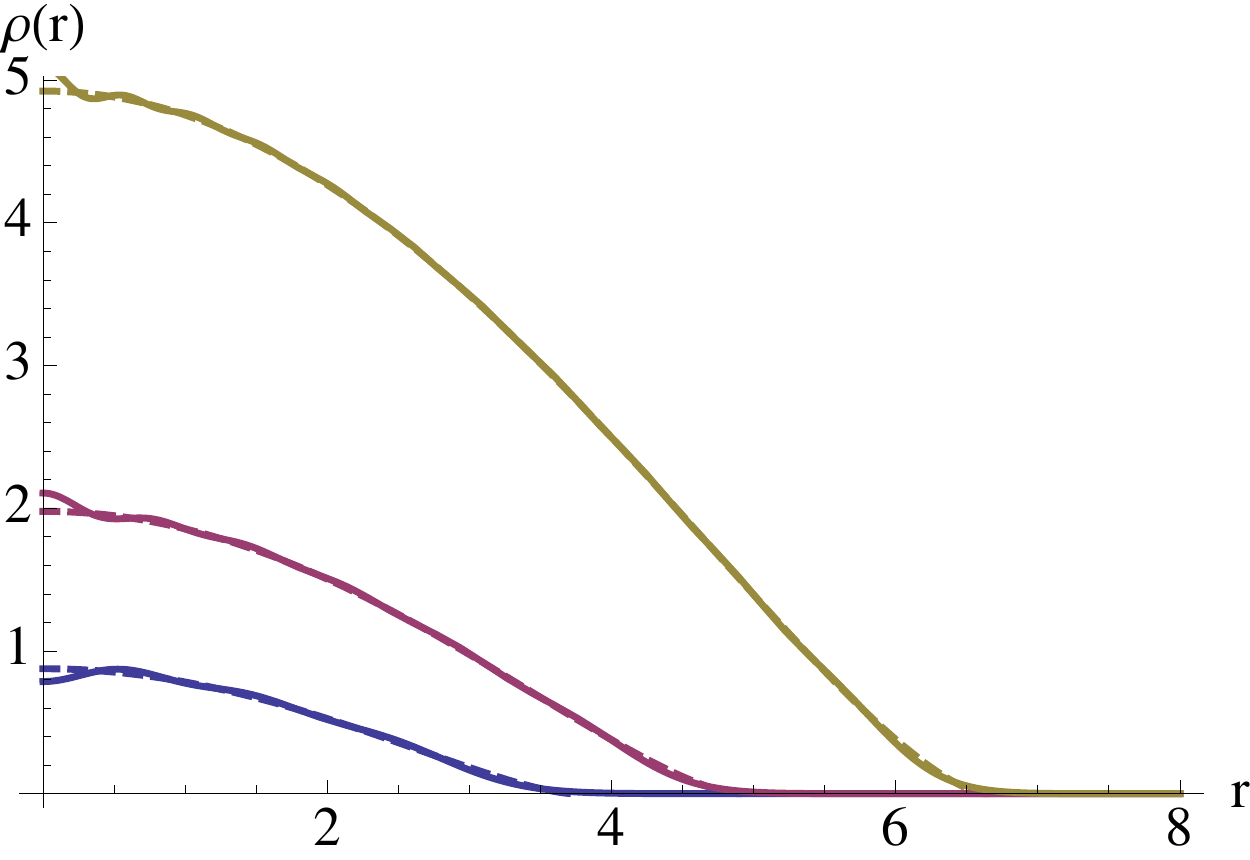}
	}
	\caption{
	\label{fig:1pdm}
		Left: the one-particle density matrix for $L=5$ and various $D$. Right: the electron density (solid) and its TF approximation (dashed) for various $D$. Plots for $L=5$ (blue), $L=10$ (red) and $L=20$ (yellow) .}
\end{figure*}

\section{One-particle density matrix and electron density}

The total number of electrons is
\begin{equation} 
\label{n}
	n = \frac{\Gamma\left(L+D+1\right)}{\Gamma\left(L+1\right)\Gamma\left(D+1\right)},
\end{equation}
where $\Gamma$ is the gamma function \cite{NISTbook}, and the one-particle density matrix is 
\begin{equation}
	\rho_1\left(\bm{r}_1,\bm{r}_2\right)
	= \sum_{\ell=0}^{L}
	\sum_{\ell_1+\dots+\ell_D = \ell}
	\Psi_{\scL}\left(\bm{r}_1\right)
	\Psi_{\scL}\left(\bm{r}_2\right).
\end{equation}
Introducing the relative and center-of-mass coordinates,
\begin{align}
	\bm{u} & = \bm{r}_1 - \bm{r}_2,&
	\bm{U} & = \frac{\bm{r}_1 + \bm{r}_2}{2},
\end{align}
the one-particle density matrix becomes \cite{Sondheimer51, vanZyl03b, Shea07}
\begin{multline}
\label{1pdm}
	\rho_1\left(u,U\right)
	= \frac{1}{\pi^{D/2}} 
	\sum_{\ell=0}^{L} (-1)^{\ell}
	\\ \times 
	L_{L-\ell}^{\frac{D}{2}}\left(\frac{u^2}{2}\right)e^{-\frac{u^2}{4}}
	L_{\ell}^{\frac{D}{2}-1}\left(2 U^2\right)e^{-U^2},
\end{multline}
where $L_{\ell}^{\lambda}$ is an associated Laguerre polynomial \cite{NISTbook}. Equation \eqref{1pdm} is derived using the connection between the inverse Laplace transform of the Bloch density matrix and the one-particle density matrix \cite{Shea07}. The one-particle density matrix is represented in Fig.~\ref{fig:1pdm} for $L=5$ and various $D$ .

The electron density can be easily obtained from \eqref{1pdm} and reads
\begin{multline}
\label{rho}
	\rho \left(r\right)
	= \frac{1}{\pi^{D/2}} 
	\sum_{\ell=0}^{L} (-1)^{\ell}
	\\ \times 
	\frac{\Gamma\left(L-\ell+\frac{D}{2}+1\right)}
	{\Gamma\left(L-\ell+1\right)\Gamma\left(\frac{D}{2}+1\right)}
	L_{\ell}^{\frac{D}{2}-1}\left(2r^2\right)e^{-r^2}.
\end{multline}

Within the Thomas-Fermi (TF) approximation \cite{Thomas27, Fermi26}, Eq.~\eqref{rho} becomes \cite{Vignolo00, Brack01, Mueller04}
\begin{equation}
\label{rho-TF}
	\rho_{\rm TF} \left(r\right) =
	\frac{\left(R_{\rm TF}^2 - r^2\right)^{D/2}}{2^{D} \pi^{D/2} \Gamma\left(\frac{D}{2}+1\right)},
\end{equation}
where
\begin{equation}
\label{R-TF}
	R_{\rm TF}^2 = 2 
	\left[\frac{\Gamma\left(L+D+1\right)}
	{\Gamma\left(L+1\right)}\right]^{1/D}
\end{equation}
measures the radial extent of the density within the TF approximation.  Fig.~\ref{fig:1pdm} shows $\rho$ and $\rho_{\rm TF}$ for various $D$ and $L$ and reveals that the TF approximation is remarkably good, even when $L$ is quite small.  It fails, however, to reproduce the fine structure that results from statistical Fermi correlations and we note that this fine structure is most pronounced when $D$ is small.

\begin{table}
\caption{
\label{tab:coeffs}
Thomas-Fermi and Dirac coefficients of the harmonically-trapped electron gas for various $D$ in the thermodynamic (large-$L$) limit.}
\begin{ruledtabular}
\begin{tabular}{cccccc}
Coefficient		&	$D=1$		&	$D=2$		&	$D=3$	
\\
\hline
$C_{\rm T}$		&	$\pi^2/6$		&	$\pi$			&	$(9\pi/5)\left(\pi/6\right)^{1/3}$
\\
$-C_{\rm X}$		& 	$1/2$	&	$8/(3\sqrt{\pi})$		&	$(3/4)\left(6/\pi\right)^{1/3}$	
\\
\end{tabular}
\end{ruledtabular}
\end{table}

\section{Kinetic and exchange energies}

The kinetic energy of the system can be easily obtained via the one-particle density matrix, and it reads
\begin{equation}
\label{ET}
\begin{split}
	E_\text{T}\left(D,L\right)
	& = -\ha \int \left. \nabla_{u}^2\,
	\rho_1\left(u,U\right)\right|_{u=0} d\bm{U}\\
	& = \frac{D}{2} \left(L+\frac{D+1}{2}\right)
	\frac{\Gamma\left(L+D+1\right)}
	{\Gamma\left(D+2\right)\Gamma\left(L+1\right)},
\end{split}
\end{equation}
which behaves as
\begin{equation}
\label{ET-largeL}
	E_\text{T}\left(D\right) \rightarrow \frac{D/2}{(D+1)!} L^{D+1}
\end{equation}
for large $L$.

Moreover, it can be shown that its exchange energy is
\begin{equation}
\label{EX}
\begin{split}
	E_\text{X}\left(D,L\right)
	& = -\ha \iint
	\frac{\rho_1\left(u,U\right)^2}{u}
	d\bm{u}\,d\bm{U}\\
	& = 
	-\frac{\Gamma\left(\frac{D-1}{2}\right)}
	{\sqrt{2\pi}\,\Gamma\left(D/2+1\right)
	\Gamma\left(D/2\right)^2}
	\sum_{\ell=0}^{L} \\
	& \times 
	\frac{\Gamma\left(\frac{D}{2}+\ell\right)
	\Gamma\left(L-\ell+\tha\right) 
	\Gamma\left(\frac{D}{2}+1-\ell+L\right)}
	{\Gamma\left(\ell+1\right)
	\Gamma\left(L-\ell+1\right)^2}\\
	& \times {}_3F_2
	\left[
	\begin{array}{c}
	\begin{array}{ccc}
	-1/2,		&\frac{D-1}{2},	&	\ell-L\\
	\end{array}\\
	\begin{array}{cc}
	D/2+1,		&\ell-L-\ha
	\end{array}
	\end{array}
	;1
	\right],
\end{split}
\end{equation}
where ${}_3F_2$ is the generalized hypergeometric function. One can verify that, for $D=2$, Eq.~\eqref{EX} reduces to the expression given in Ref.~\cite{vanZyl03b}. In the limit of large $L$, it becomes
\begin{equation}
\label{EX-largeL}
	E_\text{X}\left(D\right) \rightarrow - \frac{\sqrt{2}}{\pi} \frac{\Gamma\left(\frac{D-1}{2}\right)}{\Gamma\left(\frac{D}{2}\right)\Gamma\left(D+\frac{3}{2}\right)} L^{D+1/2}.
\end{equation}

\section{Thomas-Fermi and Dirac coefficients}

The kinetic and exchange energies can also be obtained using the TF \cite{Fermi26, Thomas27} and Dirac \cite{Dirac30} functionals, which read
\begin{gather}
	E_\text{T}(D,L) = C_{\rm T}(D,L) \int \rho(r)^{1+2/D} d\bm{r},
	\label{ET-TF}\\
	E_\text{X}(D,L) = C_{\rm X}(D,L) \int \rho(r)^{1+1/D} d\bm{r}.
	\label{EX-TFD}
\end{gather}

In the thermodynamic (large-$L$) limit, $\rho(r)$ can be replaced by $\rho_{\text{TF}}(r)$, and, after integration, we have
\begin{gather}
	E_\text{T}(D) = C_{\rm T}(D) \frac{1}{4\pi} \frac{D+2}{D+1} \frac{\Gamma(D+1)^{1/D}}{\Gamma\left(\frac{D}{2}+1\right)^{2/D}} n^{1+1/D},
	\label{ET-TF-int}\\
	E_\text{X}(D) = C_{\rm X}(D)  \frac{1}{\sqrt{2\pi}} \frac{\Gamma(\frac{D+3}{2})}{\Gamma(D+\frac{3}{2})} \frac{\Gamma(D+1)^{1+\frac{1}{2D}}}{\Gamma\left(\frac{D}{2}+1\right)^{1+\frac{1}{D}}} n^{1+\frac{1}{2D}}.
	\label{EX-TFD-int}
\end{gather}
Equating \eqref{ET-largeL} and \eqref{EX-largeL} with \eqref{ET-TF-int} and \eqref{EX-TFD-int} yields
\begin{gather}
	C_{\text{T}}(D) = 2\pi\frac{D}{D+2} \Gamma\left(\frac{D}{2}+1\right)^{2/D},
	\label{CT}
	\\
	C_{\text{X}}(D) = -\frac{4}{\sqrt{\pi}}\frac{D}{D^2-1}
	\Gamma\left(\frac{D}{2}+1\right)^{1/D}.
	\label{CX}
\end{gather}
 For $D=1$, the exchange energy is infinite because it must compensate the Coulomb energy, which is also infinite \cite{Lee11, 2QR}. However, one can determine the value of the coefficient $C_\text{X}$(1) by replacing the Coulomb interaction by a short-ranged interaction potential (see Appendix \ref{app:Dirac}). The resulting values of $C_\text{T}$ and $C_\text{X}$, which are gathered in Table \ref{tab:coeffs}, are identical to the $D$-jellium expressions \cite{Vignale}, showing that, in the thermodynamic limit, the two paradigms are equivalent.

\begin{figure*}
	\subfigure[~$D=1$]{
	\includegraphics[width=0.45\textwidth]{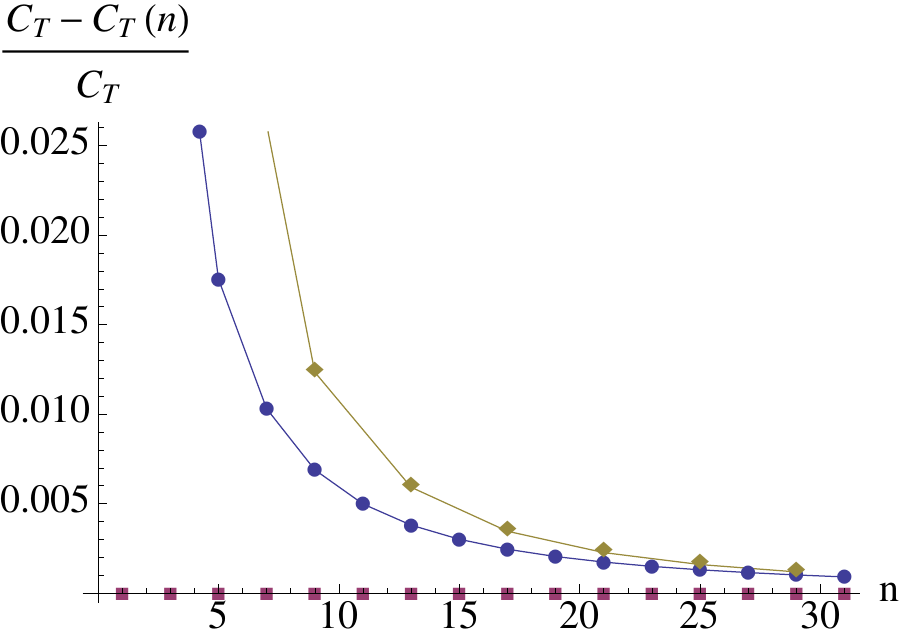}}
	\\
	\subfigure[~$D=2$]{
	\includegraphics[width=0.45\textwidth]{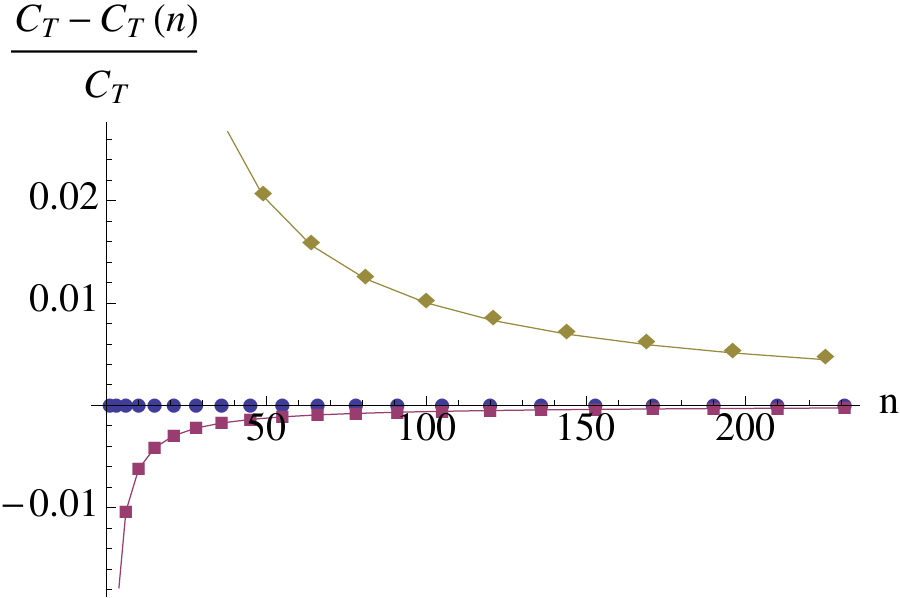}
	\includegraphics[width=0.45\textwidth]{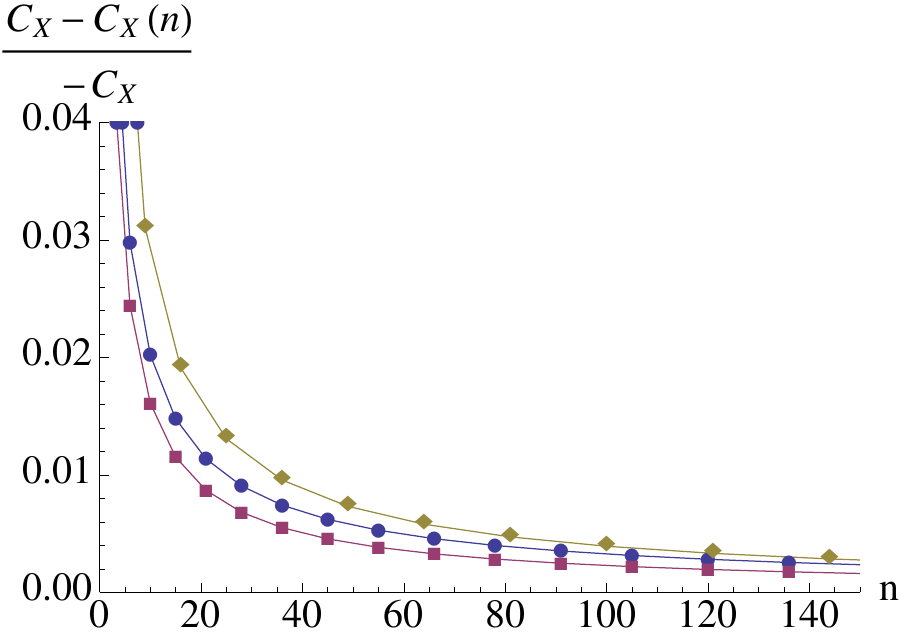}}
	\subfigure[~$D=3$]{
	\includegraphics[width=0.45\textwidth]{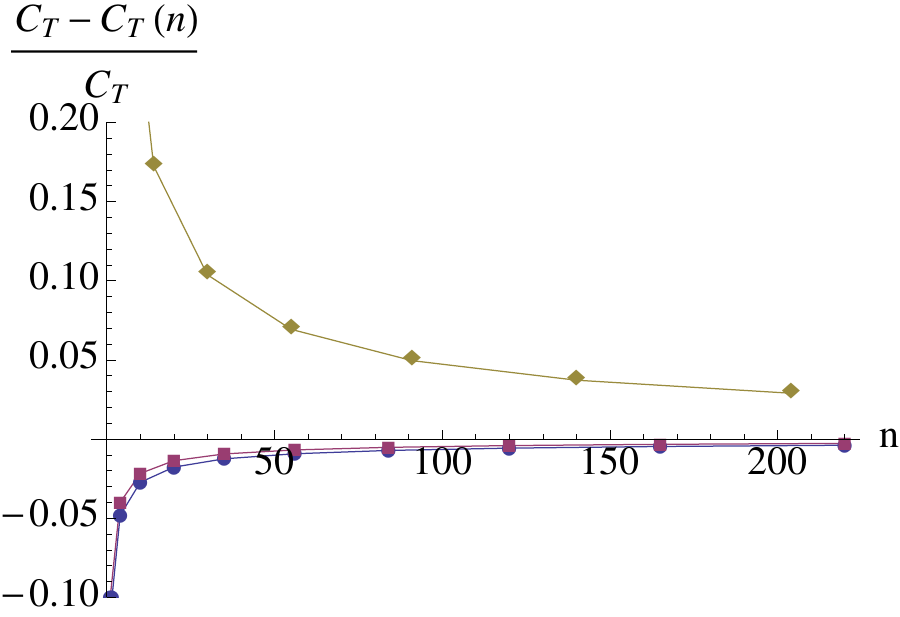}
	\includegraphics[width=0.45\textwidth]{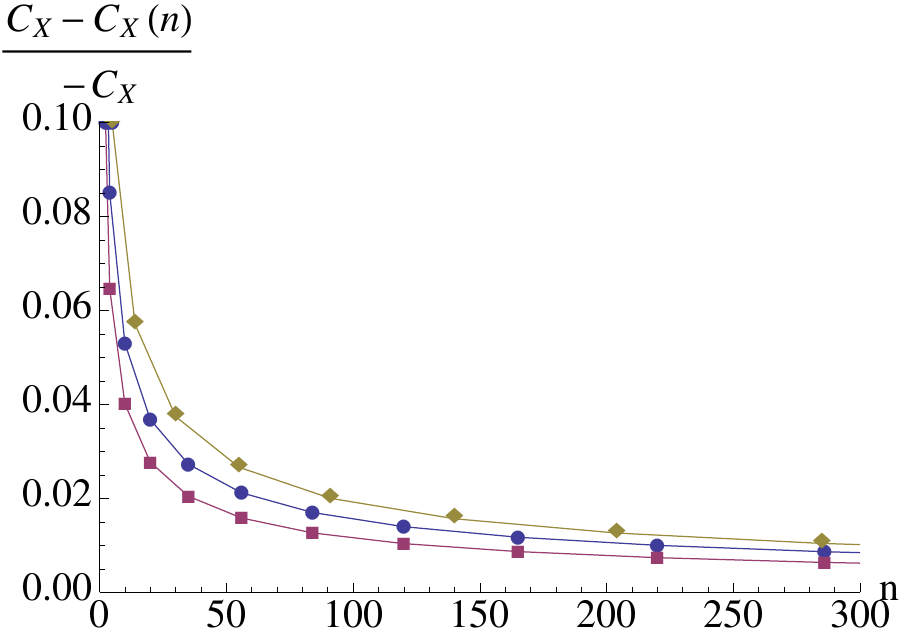}}
	\caption{
	\label{fig:CT-CX}
	Convergence of the Thomas-Fermi and Dirac coefficients for various $D$ with respect to the number of electrons $n$ for the harmonically-trapped jellium model using the true density $\rho(r)$ (blue circles) and the TF density $\rho_{\text{TF}}(r)$ (red squares) and the spherium model (yellow diamonds).
	}
\end{figure*}

Several observations can be made from Fig.~\ref{fig:CT-CX}, which shows how the coefficients $C_\text{T}$ and $C_\text{X}$ evolve with the number of electrons $n$ for various $D$.  For $D=1$, one sees that the values of $C_\text{T}$ in spherium are different from those in the harmonic jellium model, but follow the same trend. For $D=2$, it turns out that the TF functional \eqref{ET-TF} is actually exact for the harmonic jellium model \cite{Brack01}. In other words, it means that, using the exact non-interacting electron density $\rho(r)$, one can get the exact value of the non-interacting kinetic energy (no gradient correction is needed). 

This applies to the one-dimensional case if one uses the TF density instead of the true density.  We note that for both $D=2$ and $D=3$, the spherium values for $C_\text{T}$ follow different trends from the harmonic jellium model, but converge to the same limiting values.  For the $C_\text{X}$ coefficient, one finds that the harmonic jellium and spherium values are similar, and it may be possible to use the closed-form expressions of the $C_\text{X}$ coefficient in spherium to estimate the exchange energy in non-uniform systems \cite{Glomium, UEGs}.

\section{Conclusion}

In this article, we have studied the non-interacting kinetic and exchange energies for a system consisting of $n$ electrons trapped in an isotropic harmonic potential. We have shown that, in the thermodynamic limit, this paradigm is identical to the conventional uniform electron gas (jellium) and the spherium model. Particular attention has been devoted to the study of the convergence of the Thomas-Fermi and Dirac coefficients as functions of the number of electrons for various values of the dimensionality. We hope that our results will be useful to understand finite-size effects in homogenous and inhomogeneous systems within DFT.

\begin{acknowledgments}
We thank Joshua Hollett for useful discussions, the NCI National Facility for a generous grant of supercomputer time and the Australian Research Council (Grants DP0984806, DP1094170 and DP120104740) for funding.
\end{acknowledgments}

\appendix

\section{\label{app:Dirac}
Dirac coefficient for $D=1$}

The coefficient $C_{\rm X}$ for $D=1$ can be found by replacing the Coulomb operator by a short-ranged interaction potential
\begin{equation}
	\frac{1}{\left|x_1-x_2\right|} \rightarrow \delta\left(x_1-x_2\right),
\end{equation}
where $\delta$ is the Dirac delta function. This is commonly done in the literature \cite{Yang67, Sutherland68, Fogler05} due to the divergence of the Coulomb operator for small interelectronic distances in one dimension.

It follows that
\begin{multline}
\label{EX-1D}
	E_{\rm X}\left(1,L\right) 
	= -\frac{1}{2\sqrt{2\pi}}
	\sum _{\ell_1,\ell_2=0}^{L}
	(-1)^{\ell_2}\frac{\Gamma\left(\ell_1-\ell_2+\ha\right)}
	{\Gamma(\ell_2+1)}\\
	\times {}_3\Tilde{F}_2
	\left[
	\begin{array}{c}
	\begin{array}{ccc}
	\ha,		&\ell_1-\ell_2+\ha,	&	-\ell_2\\
	\end{array}\\
	\begin{array}{cc}
	\ha-\ell_2,	&\ell_1-\ell_2+1
	\end{array}
	\end{array}
	;1
	\right]
\end{multline}
where ${}_3\Tilde{F}_2$ is a regularized hypergeometric function \cite{NISTbook}. Equation \eqref{EX-1D} yields 
\begin{equation}
	C_{\rm X}(1) = -\frac{1}{2},
\end{equation}
which is identical to the one-dimensional jellium \cite{Vignale} and spherium \cite{Unpub} systems.

\end{document}